\documentclass[%
superscriptaddress,
 amsmath,amssymb,
twocolumn,
 aps,
prb,
]{revtex4-2}

\usepackage{graphicx}
\usepackage{dcolumn}
\usepackage{bm}
\usepackage{lineno}
\usepackage[colorinlistoftodos]{todonotes}
\usepackage{xcolor}
\usepackage{mathtools}
\usepackage{array}
\usepackage{textcomp}
\newcommand{\defeq}{\mathrel{\mathop:}=}

\makeatletter
\def\fm@sep{}
\def\@author@join@#1#2#3{\def\@author{{#1}{#2\fm@sepmark\frontmatter@footnote{#3}}}}
\def\fm@sepmark{\fm@sep\def\fm@sep{\textsuperscript{,}}}
\def\doauthor#1#2#3{%
  \ignorespaces#1\unskip\@listcomma
  \begingroup
   #3%
  \@if@empty{#2}{\endgroup{}{}}{\endgroup{\comma@space}{}\begingroup\def\fm@sep{}#2\endgroup}%
  \space \@listand
}%
\makeatother

\begin{document}

\title{Nature of exciton-lattice-spin coupling in the van der Waals antiferromagnet CrSBr}

\author{Luca Nessi}
\altaffiliation{These authors contributed equally to this work.}
\email{luca.nessi@polimi.it}
\affiliation{Department of Physics, Politecnico di Milano, Milano 20133, Italy}

\author{Francesco Gucci}
\altaffiliation{These authors contributed equally to this work.}
\affiliation{Department of Physics, Politecnico di Milano, Milano 20133, Italy}
\affiliation{Department of Physics, Regensburg Center for Ultrafast Nanoscopy (RUN) and Halle-Berlin-Regensburg Cluster of Excellence CCE, University of Regensburg, Regensburg, Germany}

\author{Serena Bragadini}
\affiliation{Department of Applied Physics and Materials Science, and Department of Physics, California Institute of Technology, Pasadena 91125, USA}

\author{Khoa B. Le}
\affiliation{Department of Applied Physics and Materials Science, and Department of Physics, California Institute of Technology, Pasadena 91125, USA}

\author{Connor A. Occhialini}
\affiliation{Department of Physics, Massachusetts Institute of Technology, Cambridge 02139, USA}

\author{Andrea Iudica}
\affiliation{Department of Physics, Politecnico di Milano, Milano 20133, Italy}

\author{Armando Genco}
\affiliation{Department of Physics, Politecnico di Milano, Milano 20133, Italy}
\affiliation{Department of Physics, Università di Pisa, Pisa 56127, Italy}

\author{Ettore Carpene}
\affiliation{CNR-IFN, Institute for Photonics and Nanotechnologies, Milano 20133, Italy}

\author{Rupert Huber}
\affiliation{Department of Physics, Regensburg Center for Ultrafast Nanoscopy (RUN) and Halle-Berlin-Regensburg Cluster of Excellence CCE, University of Regensburg, Regensburg, Germany}

\author{Riccardo Comin}
\affiliation{Department of Physics, Massachusetts Institute of Technology, Cambridge 02139, USA}

\author{Marco Bernardi}
\affiliation{Department of Applied Physics and Materials Science, and Department of Physics, California Institute of Technology, Pasadena 91125, USA}

\author{Stefano Dal Conte}
\affiliation{Department of Physics, Politecnico di Milano, Milano 20133, Italy}

\author{Giulio Cerullo}
\email{giulio.cerullo@polimi.it}
\affiliation{Department of Physics, Politecnico di Milano, Milano 20133, Italy}
\affiliation{CNR-IFN, Institute for Photonics and Nanotechnologies, Milano 20133, Italy}

\date{\today}

\begin{abstract} 

\textbf{The van der Waals antiferromagnet CrSBr has emerged as a prototypical platform to explore the interplay between excitons, magnons and lattice. In the presence of a canting magnetic field, the Cr-localized exciton X$_1$ acts as an optical probe of the magnetization precession, directly linking excitonic and magnetic degrees of freedom. Here, we extend this picture to the second exciton, X$_2$, by performing broadband transient reflectance spectroscopy on a bulk crystal. Resonant excitation shows that the photoexcited exciton population launches the spin waves. Tracking the central energies of both resonances, we find that the more delocalized X$_2$ is sensitive not only to the spin dynamics, but also to structural deformations. The ratio between the magnon-induced shifts of X$_2$ and X$_1$ exceeds its static, field-induced counterpart by up to one order of magnitude, revealing an additional lattice-mediated mechanism, periodically driven by the magnons, beyond the purely electronic picture established so far. First-principle calculations show that the out-of-plane hybridization of X$_2$ makes it a sensitive probe of the modulated interlayer distance. Our results clarify the origin of optically launched magnons in CrSBr and reveal a time-dependent exciton-spin-lattice coupling, opening new perspectives for magneto-optomechanics in van der Waals magnets.}

\end{abstract}
\maketitle

\section{Introduction}
Spin-wave modes and bound excitons in magnetic semiconductors are typically decoupled, making the optical excitation of magnons highly inefficient. Conventional excitation strategies therefore exploit their interaction with other degrees of freedom, typically through electronic, structural, or magnetic channels. For example, the inverse Faraday and inverse Cotton-Mouton effects \cite{kimel2005ultrafast, kalashnikova2007impulsive} excite spin waves through virtual electronic transitions. Alternatively, magnetic order can be altered exciting specific \textit{d-d} transitions in correlated systems \cite{kampfrath_coherent_2011, schlauderer_temporal_2019, mikhaylovskiy2020resonant, afanasiev2021controlling, seifert2022ultrafast}, leveraging magnon-phonon hybridization, \cite{nova2017effective, afanasiev2021ultrafast, cui2023chirality, ilyas2024terahertz} or directly coupling to magnons or electromagnons \cite{kubacka2014large}. Conversely, spin-wave detection typically relies on magneto-optical effects, such as Faraday (Kerr) rotation or monitoring the electromagnetic emission of magnons \cite{belvin2021exciton, allington2025distinct}.

\begin{figure*}[t]
\centering
\includegraphics[width = \textwidth]{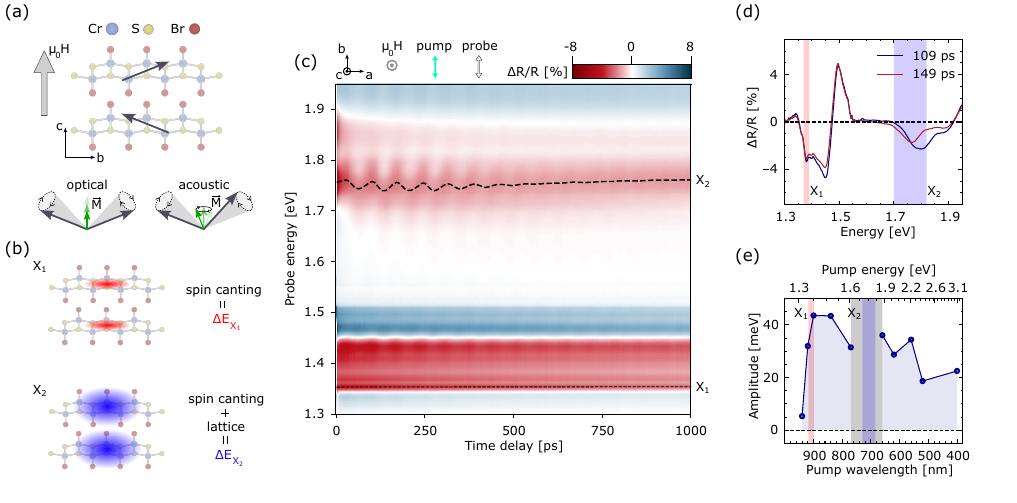}
\caption{(a) Schematic of the crystal structure and magnon modes of CrSBr. An external magnetic field of $\sim 230$ mT is applied parallel to the $\hat{c}$ axis, canting the spins (dark gray arrows) in the out-of-plane direction. At the bottom, optical and acoustic magnon modes as defined in the current configuration. (b) Representation of the Cr-site localized exciton X$_1$ (red, top) and of the more delocalized exciton X$_2$ (blue, bottom). (c) Transient reflectivity map up to 1 ns with a pump pulse of energy 2.21 eV and fluence 300 \textmu J/cm$^2$. X$_1$ and X$_2$ peaks are tracked using dashed black lines. (d) Transient spectra at selected delays that maximize the relative shift of the excitonic peaks. The shaded red and blue areas highlight the X$_1$ and X$_2$ regions, respectively. (e) Pump energy dependence at fixed fluence of the magnon-mediated oscillation amplitude of X$_2$ energy ($\Delta\mathrm{E_{X_2}}$). All the measurements were performed at 77 K and with an external magnetic field aligned along $\hat{c}$ of $\sim 230$ mT.}
\label{fig1}
\end{figure*}

Ref. \cite{bae2022exciton} has shown that in CrSBr -- a van der Waals A-type antiferromagnetic semiconductor -- magnons couple to excitons via interlayer electronic coupling \cite{wilson2021interlayer, liebich_controlling_2025}. This mechanism enabled the observation of spin wave oscillations by tracking the optical reflectivity at the exciton energy. 

In particular, the magnon-mediated modulation of the magnetization renormalizes the exciton delocalization, modifying its central energy. Such effect has been extensively studied for the Cr-site localized exciton X$_1$ in the NIR region, around 1.36 eV. Conversely, time- and energy-resolved studies of X$_2$ (centered at 1.75 eV) are still missing. Oppositely to the Frenkel-type exciton X$_1$, the higher energy feature X$_2$ is an interband exciton with a Wannier character.
The magnetization-mediated energy renormalization has also been observed in the static regime where the spin canting, obtained by applying an external magnetic field along the $\hat{a}$ or $\hat{c}$ axis, causes an energy shift for the lowest energy exciton X$_1$, following $|\Delta\mathrm{E_{X_1}| \propto M^2}$ (where $M$ is the magnetization), up to magnetization saturation \cite{wilson2021interlayer, heissenbuttel2025quadratic}. Interestingly, the higher energy exciton X$_2$ undergoes the same effect, with a five-fold increase of the shift, associated to its more delocalized nature which enhances its sensitivity to the interlayer hopping \cite{nessi2024magnetic, datta2025magnon}.

More recently, studies of magnon propagation \cite{bae2024transient, sun2024dipolar} and their nonlinear properties \cite{diederich2023tunable,diederich2025exciton} assigned the magnon generation mechanism to magnetoelastic coupling \cite{bae2024transient, fei2024spin}. A short light pulse can trigger a magnetic response by inducing lattice oscillations, particularly a time-varying transient strain field induced by out-of-plane transverse and in-plane longitudinal lattice displacements. This was indirectly inferred by time-resolved Kerr rotation \cite{bae2024transient, fei2024spin}. Under a near-infrared (NIR) pulsed excitation, a coherent acoustic phonon (CAP) mode has been revealed by monitoring the oscillation of Bragg peaks in time-resolved electron diffraction \cite{ranhili2025ultrafast} and has recently been confirmed using optical probes \cite{shen2025orthogonal, shen2026electron}.

Beside acting as a passive probe of the instantaneous magnetization, the exciton X$_1$ also plays an active role by exerting a spin torque \cite{brennan2026excitonic}. This alters the magnetic potential energy surface, providing a tunable knob to control magnetization dynamics. Interestingly, also the higher-energy exciton X$_2$ was shown to have a non-trivial interaction with magnetic orderw. Specifically, Ref. \cite{datta2025magnon} showed that the exciton-exciton (X$_2$-X$_2$) attraction, mediated by the interlayer hybridization, affects the magnetization direction. It was further shown that the measured fluence-dependent energy shift is directly related to an additional spin canting. 

These findings highlight the rich interplay between excitons, lattice and magnons, motivating further investigations of their coupled dynamics. In particular, time-resolving their interaction offers a unique opportunity to gain deeper insight into the coupling between the optical, structural and magnetic properties, as well as the mechanisms underlying the optical generation of magnetic excitations. So far, the transient optical response of CrSBr has been investigated primarily through single-color experiments, either in the visible spectral range or targeting the NIR region around its lowest-energy exciton.

In this work, we use broadband ultrafast transient reflectivity at low temperature ($T<T_{\mathrm{N}}$) to study the dynamics of the two excitonic species $X_1$ and X$_2$ in CrSBr and their coupling to magnetic excitations. We quantify the magnon-induced energy shifts by tracking the X$_1$ and X$_2$ peaks in the dynamic reflectivity spectra.
By varying the pump photon energy, we experimentally demonstrate  that the excitonic population is the main driver of coherent magnetic precession. 
As previously stated, in the static regime, both X$_1$ and X$_2$ exhibit magnetic-field-dependent energy shifts ($|\Delta \mathrm{E_{X_1, X_2}}| \propto \mathrm{M}^2$), with the shift of X$_2$ approximately five times larger than that of X$_1$ \cite{nessi2024magnetic, datta2025magnon}. By quantitatively tracking the dynamics of both exciton energies, we find that the amplitude of $\Delta E_{\mathrm{X_2}}$ is influenced by mechanisms beyond magnetic precession, differently from X$_1$. We assign this additional contribution  to exciton-magnon coupling to a lattice-driven mechanism.

Using first-principle Bethe-Salpeter equation (BSE) for different interlayer distances, we show that the X$_2$ energy shifts with a much stronger dependence on interlayer distance than X$_1$. We attribute this greater sensitivity to the more delocalized nature of X$_2$ \cite{nessi2024magnetic, smiertka2026distinct}, which responds not only to magnetization dynamics, but also to dynamic variations of the interlayer distance. This structural sensitivity is further supported by the observation of a low-frequency vibrational mode that couples selectively to X$_2$.

Taken together, these results clarify how magnonic and phononic excitations shape the optical response of CrSBr, highlighting the importance of exciton-spin-lattice coupling in van der Waals magnetic semiconductors. The exciton-magnon-phonon coupling revealed here expands the landscape of opto-mechanic-spintronics.

\section{Experimental results}
We perform ultrafast transient reflectivity experiments on high quality CrSBr bulk crystals using a tunable narrowband pump pulse and a broadband probe pulse (see also Supplemental Material (SM) \cite{suppmat}). The crystals are cleaved under ambient conditions to expose a fresh, atomically flat surface with clear macroscopic edges, enabling identification of the two principal axes $\hat{a}$ and $\hat{b}$ (see SM \cite{suppmat}). The crystals are then mounted in a cryostat and the measurements are performed at $T=30-77$ K below the Néel temperature $T_{\mathrm{N}} \simeq 132$ K \cite{lopez2022dynamic}, while the pump and probe polarizations are aligned parallel either to the $\hat{b}$ or to the $\hat{a}$ axis of the CrSBr crystal. A small external magnetic field of a few hundreds mT is applied to cant the spins either along the out-of-plane direction $\hat{c}$ or in the $ab$ plane. For a fixed external magnetic field, two magnon branches of opposite parity are present, namely the optical and acoustic modes (depicted in Fig. \ref{fig1}(a) for \textmu$_0$$ H\parallel\hat{c}$). The former induces a modulation of the magnetization along $\hat{c}$, which is sensed by the excitons and consequently leads to an energy shift. The latter, on the other hand, drives a precession of the magnetization without affecting the excitonic energy. Small deviations of the external magnetic field from the principal axis $\hat{c}$ make also this second mode bright \cite{diederich2023tunable}.

\begin{figure*}[t]
\centering
\includegraphics[width = \textwidth]{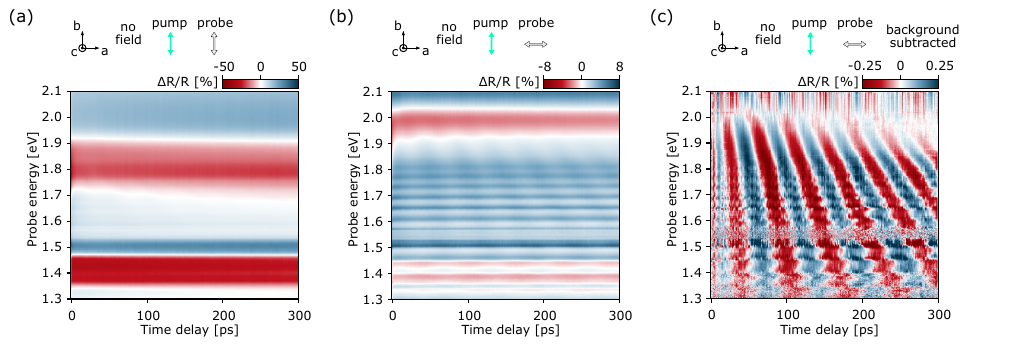}
\caption{(a) Transient reflectivity map up to 300 ps with a pump pulse of energy 3.1 eV and fluence 2.7 mJ/cm$^2$, with pump and probe pulses polarized along the $\hat{b}$ axis. (b) Same as (a) but with the probe polarized along the $\hat{a}$ axis. (c) Transient map of panel (b) without the incoherent background. All the measurements were performed at 77 K without any external applied magnetic field.}
\label{fig2}
\end{figure*}

In concert with previous studies on CrSBr, for light polarized along the $\hat{b}$ axis, we label the peak in the NIR ($\sim 1.35$ eV) as X$_1$ and the one in the visible range ($\sim 1.75$ eV) as X$_2$ \cite{nessi2024magnetic} (see SM \cite{suppmat}). The two excitonic species are characterized by distinct spatial distributions of their wavefunctions [see Fig.\ref{fig1}(b)] \cite{smiertka2026distinct}. X$_1$ exhibits a more pronounced Frenkel character, with its wavefunction predominantly localized on Cr atoms. By contrast, X$_2$ is more delocalized and has a stronger contribution from d-p hybridized states \cite{nessi2024magnetic, datta2025magnon, smiertka2026distinct}, making it intrinsically more sensitive to lattice distortion.

\subsection{Exciton-magnon coupling and its generation mechanism}
While the static spectrum of CrSBr has been investigated extensively in previous works \cite{wilson2021interlayer, nessi2024magnetic, smiertka2026distinct}, measurements of its transient response have so far been limited to single-color time traces \cite{meineke2024ultrafast} or to the NIR region around X$_1$ \cite{bae2022exciton, bork2025excitonic, brennan2026excitonic}, precluding detailed understanding of X$_2$. We first acquire a broadband transient reflectivity map probing the range between 1.3 and 2.0 eV and collecting the response up to 1 ns at 77 K with an out-of-plane magnetic field ($\parallel\hat{c}$) of $\sim$ 230 mT. The data are reported in Fig. \ref{fig1}(c) and show distinct oscillations in the NIR and visible range, which can be attributed to coherent magnons. A Fourier analysis finds two distinct magnon modes (acoustic and optical, at $\sim$ 14 and 22 GHz), which disappear when the temperature approaches $T_{\mathrm{N}}$ as it was also demonstrated for the case of X$_1$ (full temperature dependence is reported in the SM \cite{suppmat}). In the NIR region, we clearly identify a sharp feature, associated with X$_1$ (see the dashed line at $\sim$ 1.35 eV in Fig. \ref{fig1}(c) and red shaded region in Fig. \ref{fig1}(d)), followed by a broad band up to 1.45 eV. This band originates from exciton-polaritons, whose formation is enabled by the high refractive index of CrSBr, and it is a characteristic feature of thick bulk CrSBr samples (the role played by the sample thickness is further elaborated in SM \cite{suppmat}). In the visible region, the transient signal presents a peak around 1.75 eV associated to X$_2$ and an additional feature at $\sim$ 1.9 eV, assigned to the band gap \cite{watson2024giant, smolenski2025large}. As demonstrated in Ref. \cite{bae2022exciton}, magnons couple to excitons by renormalizing their resonances, giving rise to the strong reflectivity oscillations observed in Fig. \ref{fig1}(c). We focus on the transient signals of X$_1$ and X$_2$ and we track the position of both peaks as a function of pump-probe delay \cite{suppmat}. The broad spectral content of the white-light probe allows us to directly track the position of the two peaks simultaneously, enabling a direct comparison between their dynamics. The extracted time traces are superimposed onto the map of Fig. \ref{fig1}(c) with dashed black lines.

Figure \ref{fig1}(d) shows two transient spectra, obtained by taking vertical cuts of the pump-probe map in Fig. \ref{fig1}(c) at a crest and at a dip of the spin wave oscillation. We extract the exciton energies by fitting the time-resolved spectra with Gaussian functions. The energy shift - defined as the difference between the maximum and minimum energy over time - for X$_2$ is close to $\sim$ 20 meV, whereas for X$_1$ it is $< 0.5$ meV. In the out of equilibrium case, the ratio $r \defeq \Delta E_{\mathrm{X_2}}/\Delta E_{\mathrm{X_1}}>40$ between the energy shifts of X$_2$ and X$_1$ after excitation of coherent spin waves is substantially larger than the corresponding  $r\sim$ 5 ratio observed in the static regime by canting the antiparallel magnetic moments of each layer with an external magnetic field \cite{nessi2024magnetic}. This provides a first indication that the mechanism underlying the X$_2$ energy shift cannot be attributed solely to the instantaneous spin canting angle. The distinct transient dynamics of the X$_2$ energy compared to X$_1$ requires an additional mechanism beyond the interlayer electronic hopping proposed for X$_1$ in the static regime under an applied magnetic field. \cite{wilson2021interlayer, bae2022exciton, brennan2026excitonic}.

We next record transient reflectivity maps while changing the pump photon energy and track the coherent response of X$_2$. The maximum amplitude of the X$_2$ energy oscillations in the transient response, as a function of the pump photon energy, is shown in Fig. \ref{fig1}(d) for a constant pump fluence of 300 \textmu J/cm$^2$. As the pump photon energy is scanned from high to low values, the oscillation amplitude does not simply follow the CrSBr absorption spectrum \cite{suppmat}.  Rather, it increases as the pump is tuned across the X$_1$ ($\sim$ 1.35 eV) resonance, reaching a maximum of $\sim 40$ meV. Upon further detuning to lower energies, the amplitude drops sharply to $\sim 5$ meV. These results directly demonstrate that the X$_1$ population itself is the dominant mechanism responsible for launching magnons under optical pumping. This evidence is further corroborated by noting that pumping in a transparency region does not by itself imply the absence of magnetic excitation: oscillations with amplitude exceeding 20 meV are recorded when pumping the spectral region between X$_1$ and X$_2$. We could not determine the X$_2$ dynamic renormalization under resonant excitation because of pump scattering artifacts [gray shaded area in Fig. \ref{fig1}(d)].

\subsection{Pump-induced lattice distortion}
To isolate the effects of structural distortion on the excitonic species, we perform additional pump-probe measurements in the absence of an external magnetic field. Figure \ref{fig2}(a) shows the results for a high pump photon energy (3.1 eV) and high fluence (2.7 mJ/cm$^2$). In agreement with the literature \cite{bae2022exciton}, we observe no clear oscillations associated with a magnetization precession, which requires an initial canting of the spins. However, we observe a faint oscillation below 1.7 eV, in the transparency region between the excitons. To further investigate this effect, we repeat the experiment with the probe polarized along the $\hat{a}$ axis [see Fig. \ref{fig2}(b)]. Here, excitons have negligible oscillator strength and CrSBr becomes transparent, allowing for longer propagation distances of the probe polarization. The spectral oscillations related to Fabry-Pérot modes can be observed and they are related to the crystal thickness, whereas oscillations are also visible in the time traces all over the probe spectrum. Figure \ref{fig2}(c) shows the transient reflectivity map presented in Fig. \ref{fig2}(b) after subtraction of the slowly varying incoherent component. The oscillations frequency depends on the probe wavelength. 
We attribute these oscillations to coherent acoustic phonons (CAPs), whereby the probe pulse is reflected by the strain wave generated by the pump pulse through a Brillouin scattering process \cite{thomsen1986surface, pontecorvo2011visualizing}. The relation between the observed CAPs frequencies and the probe wavelengths reads:
\begin{equation}
    f(\lambda_{\mathrm{p}}) = \frac{2n}{\lambda_{\mathrm{p}}}v_{\mathrm{s}}\cos\theta\text{,}
    \label{eqA1}
\end{equation}
where $n$ is the real part of the refractive index, $v_{\mathrm{s}}$ is the sound velocity, $\theta$ the scattering angle and $\lambda_{\mathrm{p}}$ is the probe wavelength. From equation \ref{eqA1}, we calculate a sound velocity associated to this strain wave of $v_{\mathrm{s}} \simeq 2183$ m/s. According to phononic band structure calculations reported in Ref. \cite{ranhili2025ultrafast}, this can be associated to the LA acoustic mode. The same mode was also observed by single-color pump probe spectroscopy in Ref. \cite{shen2025orthogonal}.

These results prove two important points. First, optical pumping induces a structural deformation caused by the deposited pump pulse energy which results in a strain wave packet that propagates from the sample surface into the bulk. Second, this propagating strain wave packet alone does not drive a coherent modulation of the X$_2$ energy [see Fig. \ref{fig2}(a)]. This lattice excitation is broad in momentum and does not carry a characteristic frequency. Its detection is due to a periodic modulation of the refractive index that couples to the probe wavelength rather than to exciton-phonon coupling. Therefore, previously reported single-color measurements that couple to a single CAP frequency might provide an incomplete picture \cite{shen2025orthogonal}. Together, these observations show that the enhanced X$_2$ energy shift in Fig.\ref{fig1}(c) is due to the interplay between structural interplay of structural and magnetic degrees of freedom, where magnons play an active role in the dynamic deformation of the lattice.

\subsection{Exciton-lattice coupling via first-principles calculations}
\begin{figure}[]
\centering
\includegraphics[width = \columnwidth]{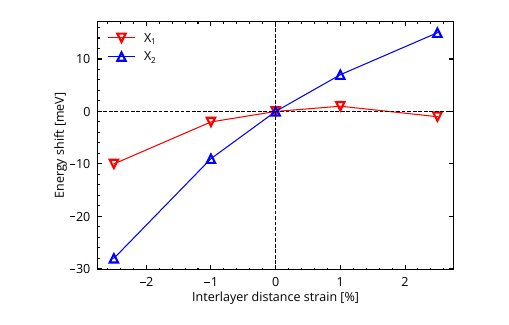}
\caption{BSE calculations of the CrSBr X$_1$ (red) and X$_2$ (blue) absorption peaks relative energies as a function of the out-of-plane interlayer distance, when applying a compressive (negative sign) and tensile strain (positive sign).}
\label{fig3}
\end{figure}

To further investigate the coupling of the lattice with the two excitonic species, X$_1$ and X$_2$, we perform first-principles Bethe-Salpeter equation (BSE) calculations on bulk CrSBr. The band structure of CrSBr in its A-type antiferromagnetic ground state is computed with DFT$+U+J$ using Quantum ESPRESSO~\cite{QE2009,QE2017}. The BSE is then solved with the Lumen fork~\cite{Lumen} of the Yambo code~\cite{Yambo2009,Yambo2019}, approximating the quasiparticle correction with a rigid scissor chosen to reproduce the measured band gap and X$_2$ energy. The absorption spectra are computed for light polarized along $\hat{b}$, for which X$_1$ and X$_2$ are observed. To assess the effect of structural deformations, we study the energy of X$_1$ and X$_2$ as a function of interlayer spacing by tracking their peaks in the calculated absorption spectrum as a function of the interlayer distance ($d_{\mathrm{IL}}$) along the ($\hat{c}$) axis. For each value of $d_{\mathrm{IL}}$, the layers are displaced rigidly along $\hat{c}$ with the in-plane lattice parameters kept fixed, the atomic positions are relaxed, and the band structure and BSE spectrum are recomputed. Further computational details are given in the Supplemental Material~\cite{suppmat}. As shown in Fig. \ref{fig3}, X$_2$ (blue triangles) exhibits a steep, approximately linear energy shift with ($d_{\mathrm{IL}}$). By contrast, the shift of X$_1$ (red triangles) is substantially smaller at strains below 2\% and has a sizable quadratic component. Near zero strain, the ratio of the slopes ($dE_{\mathrm{X}_i}/d\varepsilon$), where ($i=1,2$) and $\varepsilon$ is the interlayer strain, is approximately 5.3. The different functional forms of the energy shifts arise from the distinct trends in band-gap renormalization and exciton binding energy for X$_1$ and X$_2$, as discussed further in the Supplemental Material \cite{suppmat}. These calculations corroborate the greater sensitivity of X$_2$ to structural parameters. In the following section, we discuss magnon-lattice coupling and provide further evidence of the interplay between magnons and the lattice.

\subsection{Exciton-magnon-lattice coupling of X$_2$}
\begin{figure*}[t]
\centering
\includegraphics[width = \textwidth]{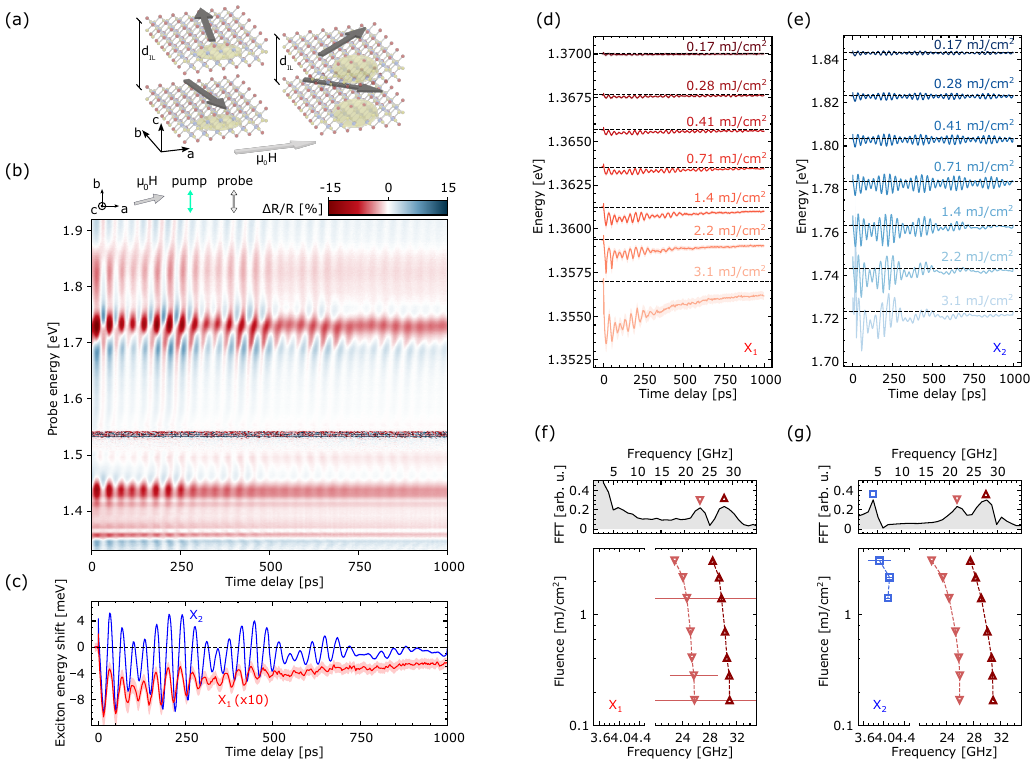}
\caption{(a) Cartoon of the physical mechanism proposed for X$_2$ energy renormalization. The interlayer distance is modulated by the generation a coherent magnon which in turn affects the exciton localization. (b) Broadband transient reflectivity map at 30 K with pump fluence of 1.4 mJ/cm$^2$. (c) Tracking of the two exciton energies X$_1$ (red line, multiplied by a factor of 10) and X$_2$ (blue). (d) Fluence dependence of the dynamical energy position from 0.17 to 3.1 mJ/cm$^2$ for X$_1$ and for (e) X$_2$. The dashed black lines indicate the equilibrium positions of X$_1$ and X$_2$ for negative delays. Data are vertically offset for clarity. (f) Fourier-transformed X$_1$ energy shift measured with a fluence of 3.1 mJ/cm$^2$ (top) and tracking of the frequencies of the main peaks appearing in the spectral response. (g) Same as (f) but for X$_2$. A low frequency mode ($\sim$ 4 GHz) appears only for large fluences.}
\label{fig4}
\end{figure*}

Our first-principles calculations suggest the following mechanisms [see Fig. \ref{fig4}(a)]. Following optical excitation of excitons, coherent spin waves are launched in the material. The magnetization precession induced by the oscillatory spin canting drives a periodic distortion of the lattice, which modulates the interlayer distance \cite{drawer2026magnon}. Simultaneously, the out-of-plane delocalization of X$_2$ is enhanced for higher tilt angles, i.e., closer to ferromagnetic configuration. The X$_2$ exciton is highly sensitive to the combined effects of these structural and magnetic changes, resulting in an enhanced amplitude of its periodic energy shift.

We show additional experiments which corroborate these observations and provide further evidence of the lattice-magnon coupling. We performed experiments at lower temperature to distinguish the X$_1$ and X$_2$ features more clearly in the reflectance contrast spectra of bulk CrSBr, enabling a more direct tracking of the peaks (see SM \cite{suppmat}). This set of experiments was performed in a pump-probe micro-spectroscopy setup featuring a closed-loop liquid helium-cooled cryostat with a base temperature of $\sim$ 30 K, further described in \cite{suppmat}. We applied an external field \textmu$_0H$ of 400 mT in the $ab$ plane, with direction tilted with respect to the $\hat{a}$ axis by $\theta_{ab}\simeq$ 15°. As elucidated in Refs. \cite{diederich2023tunable, sun2024dipolar}, introducing a tilting angle $\theta_{ab}$ couples both magnonic modes (acoustic and optical) to excitons, making them observable in transient reflectivity experiments.

Figure \ref{fig4}(b) reports a transient reflectivity map obtained for a pump fluence of 1.4 mJ/cm$^2$. The features closely resemble the ones observed in Fig. \ref{fig1}(c), with the main difference being the increased sharpness of all transitions. Moreover, the simultaneous presence of the two magnon modes with comparable weight gives rise to the strong beating in the observed oscillations. Following the approach proposed in \cite{brennan2026excitonic}, we tracked the central energy of the peaks fitting the dynamic reflectance spectra with Gaussian lineshapes (see more details in \cite{suppmat}). Figure \ref{fig4}(c) shows the energy shifts for both X$_1$ and X$_2$ with respect to the equilibrium position, which is measured at negative pump-probe delays and is depicted as a horizontal dashed line centered at zero. The two excitons undergo very different temporal dynamics. X$_1$ exhibits a fast energy renormalization of the central position, with a red-shift of $\sim$ 0.6 meV, followed by a slow recovery. Oscillations have a maximum amplitude of $\sim$ 0.5 meV. For X$_2$, the initial energy renormalization is less pronounced and the oscillations produce both a red- and a blue-shift with respect to the equilibrium position. Its maximum oscillation amplitude is $\sim$ 14 meV.

This enhancement, observed also for different experimental conditions (see Fig. \ref{fig1}), cannot be explained solely via exciton-exciton interactions\cite{datta2025magnon} because that mechanism would drive additional spin-canting, affecting both excitons simultaneously, similar to an additional applied external field. Moreover, during the magnon-mediated energy renormalization, X$_2$ undergoes a non-negligible blue-shift, whereas X$_1$ does not. In fact, the direct correspondence between spin canting angle and exciton energy recently proposed in Ref. \cite{brennan2026excitonic} cannot explain this observation highlighting a non trivial dependence of the magnetization behavior on the oscillations amplitude of $\Delta E_{\mathrm{X_2}}$.

To shed light on the time-dependent giant energy renormalization and on the large discrepancy between the behaviors of the two excitons, we tracked X$_1$ and X$_2$ energies for different pump fluences. The results are reported in Fig. \ref{fig4}(d)-(e), respectively. The qualitatively different behavior of X$_1$ and X$_2$, i.e., the ratio of the oscillations that exceeds the static counter-part and the presence of a blue-shift with respect to the equilibrium position of X$_2$, is ubiquitous for the whole fluence range.  
At low excitation fluence, the oscillations amplitude is solely modulated by the beating between the acoustic and optical magnon modes. Above $\sim$ 1 mJ/cm$^2$, on top of the beating, we observe an increasing red-shift in the central energy for X$_1$ and the appearance of a low-frequency oscillation with a period of $\sim$ 250 ps in the X$_2$ tracked position. Such slow energy renormalization is present only for X$_2$ while it is negligible for X$_1$ (see also \cite{suppmat}). Interestingly, this trend survives the decoherence of the two magnonic modes, extending beyond 1 ns. 

To gain a clearer understanding, we Fourier transform the traces reported in Fig. \ref{fig4}(d) and \ref{fig4}(e) [see Fig. \ref{fig4}(f) and \ref{fig4}(g), respectively]. The two magnon modes (light and dark red triangles) undergo the same red-shift for both excitons, from $\sim$ 26 (31) GHz at low fluence to 23 (28) GHz at high fluence. We associate this fluence-dependent shift to a photoinduced local temperature increase. At progressively higher values of the pump fluence, an additional peak at $\sim$ 4 GHz appears (blue squares) for X$_2$ only. This additional resonance matches the the long-lived slow oscillations discussed above, and we attribute it to an acoustic mode which modulates the interlayer distance.

The appearance of extra modes has already been experimentally observed and it has been attributed to nonlinear magnon effects \cite{diederich2023tunable, diederich2025exciton}. However, in our measurements the central frequency of the low-energy oscillation is not compatible with a difference frequency generation between the magnon modes: neither the magnitude nor the energy match frequency mixing processes \cite{diederich2023tunable, diederich2025exciton} (see \cite{suppmat}). Specifically, the amplitude of the low-energy peak exceeds the magnon modes, making it the predominant feature at high powers [see Fig. \ref{fig4}(g), top panel]. Ultimately, it persists longer than magnetic oscillations and it affects only X$_2$, whereas it is absent for X$_1$. Note that this observation is substantially different from what has been reported in Ref. \cite{shen2025orthogonal} (see also SM \cite{suppmat}) because in our experiment the acoustic mode couples selectively to the X$_2$ transition, resulting in an energy shift well above 5 meV.

These observations suggest that the energy modulation of  X$_2$ has contributions from different physical mechanisms and it reflects the Wannier-like character of its wavefunction. The delocalized nature of X$_2$ makes it particularly sensitive to the lattice dynamics. These results demonstrate the intimate coupling between the crystal structure and X$_2$ energetics. 

\section{Discussion}
The magnon-mediated non-equilibrium responses of the X$_1$ and X$_2$ excitons in CrSBr are dominated by different mechanisms. We established that the excitation of the X$_1$ population triggers magnon generation by performing pump photon-energy dependent measurements. Fluence dependent measurements further showed that, while the transient energy of X$_1$ follows the projection of the net magnetization precessing around its transient equilibrium position, the transient energy renormalization of X$_2$ cannot be explained solely through the hopping mechanism proposed for X$_1$ \cite{wilson2021interlayer, bae2022exciton, brennan2026excitonic}.

By comparing the relative shift of the two excitons, we showed that X$_1$ and X$_2$ differ in both their qualitative and quantitative behavior. The magnitude of the oscillations, the strong blue-shift and the presence of additional modes are not consistent with a purely electronic mechanism. We assign these observations to a magnetically-driven and lattice-mediated mechanism that together shape the energy landscape of X$_2$, making it a more sensitive probe of pump-induced modifications in bulk CrSBr. Indeed, its delocalized nature makes it more sensitive to interlayer modifications, as shown by BSE calculations, and we demonstrated that a strain wave is not by itself enough either to trigger a magnetic precession nor to couple to X$_2$ energy. Hence, the presence of a magnetic excitation makes the coupling between X$_2$ and the lattice possible.

It has been demonstrated that static uniaxial strain in CrSBr red-shifts the X$_1$ energy and this was related to an interlayer magnetic coupling that switches from AFM to FM \cite{cenker2022reversible}. Here, we went a step forward, showing how this mechanisms holds also in a dynamical framework and that the magnetic excitation triggered by the excitonic population can drive this coupling. We foresee that direct time-resolved diffraction monitoring of the dynamic response of the structure could further clarify this mechanism.

In conclusion, we probed the time-resolved dynamics of both excitonic features in bulk CrSBr following optical excitation. We experimentally demonstrated that the excitonic population is responsible for launching spin waves in this material. Furthermore, by simultaneously tracking the energies of X$_1$ and X$_2$, we showed that the energy shift of X$_2$ exceeds the value due to spin canting and the resulting interlayer electronic hopping. Our work highlights the role of X$_2$ as a sensitive probe of exciton-spin-lattice coupling, expanding the understanding of a material with a very rich physics underlying its optical spectrum.
We also showed how a magnetic excitation drives a periodic structural deformation, revealed by the enhanced energy renormalization of X$_2$ and by additional low-frequency mode coupled selectively to its resonance.
These results unify the picture of the dynamics of excitons with different characteristic size and their connection to crystal structure and magnetization. They also shed new light onto the lattice deformation driven by optically launched magnons, highlighting the importance of exciton-lattice-spin coupling in layered magnetic systems.

\textit{Acknowledgements.} The authors acknowledge Ahmet Kemal Demir, Charles Sayers, Andrea Schirato for fruitful discussions. The authors acknowledge financial support by the European Union’s EIC Pathfinder Open programme (project QUONDENSATE, 101130384).

\textit{Data availability.} Data will be available at [URL].

\bibliography{bibliography}

\end{document}